\documentstyle[times]{l-aa}

\begin{document}

   \thesaurus{}
   \title{The close DAO+dM binary RE J0720$-$318: a stratified white
dwarf with a thin H layer and a possible circumbinary disk}

   \author{M. R. Burleigh \and 
 M. A. Barstow  \and P. D. Dobbie}

   \offprints{Matt Burleigh, mbu@star.le.ac.uk}

   \institute{Department of Physics and Astronomy, University of
                Leicester \\
                Leicester LE1 7RH, UK  }

   \date{Received 13th August 1996; accepted 3rd October 1996}

   \headnote {{\it Letter to the Editor}}

     \maketitle

    \markboth{M. R. Burleigh et al.: RE
              J0720$-$318: a stratified white
             dwarf with a thin H layer and a possible 
             circumbinary disk}{RE J0720$-$318}

\begin{abstract}

We have analysed the EUVE spectrum of the DAO white dwarf RE J0720$-$318.
In contrast to the optical spectrum, which can only be fitted with a
homogeneously mixed H$+$He atmosphere, we find the EUVE spectrum can only
be matched with a stratified structure. The H layer mass of
3$\times$10$^{-14}$M$_\odot$ is a factor of $\sim$10 below upper limits
from previous EUVE observations of white dwarfs. In addition, we detect an
unprecedented HeI/HI ratio of $\sim$1 for the absorbing column along the
line of sight,  
implying an H ionization fraction $>$90\% if all this material resides in 
the local ISM. We suggest that, since this is a close pre-CV binary system,
most of the helium probably lies  
within the immediate vicinity of the star, possibly in the form of a
circumbinary disk left over from the common envelope (CE) phase. 
These results have important implications for our understanding of the
evolutionary status of DAO white dwarfs in particular, and for post-CE
binaries in general.

   \end{abstract}

%
%  14.Sep.'90: Demo-Vs.
%________________________________________________________________

\section{Introduction}

Close binary systems comprising a white dwarf plus a main sequence
companion are thought to have evolved via a phase of common envelope
evolution (e.g. Paczynski 1976), 
and are the progenitors of cataclysmic variables (e.g. de Kool 1992). Only
a few of these systems have been studied in detail, e.g. Feige 24 (DA+dM, 
Thorstensen et al. 1978) 
and V471 Tauri (DA+K2V, Nelson and Young 1970). A full understanding of the 
properties of the system and the individual stars is vital to our 
understanding of their evolutionary status and for theoretical models of
close binary evolution.

DAO white dwarfs are a small, peculiar class of degenerate objects whose 
evolution is not well understood. They have hybrid spectra showing both 
hydrogen and helium lines, in particular a weak HeII feature in the
optical at 4686\AA, and lie in the temperature range 50,000-70,000K.
Originally
it was thought that DAOs represented white dwarfs crossing between the 
helium-rich cooling sequence (DO stars) and the hydrogen-rich sequence
(DAs), arising from DOs by gravitational settling of
helium and upward diffusion of hydrogen in the atmosphere to form an overlying
hydrogen layer, leading eventually to the DAs. 
The absence of any known DAs above 70,000K and the gap in
the helium cooling sequence between 30,000-45,000K supported this idea. If this 
interpretation was correct, they should have chemically stratified
atmospheres.

Two discoveries undermined this hypothesis. Firstly, a very hot DA was
found (RE J1738+665, Barstow et al. 1994), providing a possible missing link
between 
hydrogen-rich Planetary Nebula Nucleii (PNN) and the DA cooling sequence,
and avoiding the need to invoke evolution through the helium channel first. 
Secondly, Bergeron et al. (1994) showed that almost all DAOs had homogeneously 
mixed rather than layered atmospheres. It now appears that the label `DAO'
covers  a diverse array of objects. Bergeron et al. identified at least five 
subclasses, the largest of which appears to be a group of low mass stars 
(M$<$0.48M$_\odot$) which are the direct descendants of hydrogen-rich 
subdwarfs, having evolved from the Extended Horizontal Branch (EHB).
However, at least one of these objects has been shown to be a double degenerate
(Feige 55, Holberg et al. 1995). Other subclasses include higher mass stars
consistent with post-AGB evolution, and a stratified star that could
still represent a transitional state on the cooling sequence. 
The classification of the very high mass white 
dwarf GD50 as a DAO from its EUV spectrum has further confused the picture, 
since this star may well be the result of a 
white dwarf merger (Vennes et al. 1996). The origin of the helium 
in each of these subgroups could be different. 

Excluding GD50, three DAO stars have been detected in
 the extreme ultraviolet (EUV) - RE J1016-053
(Tweedy et al. 1993), RE J2013+400 (Barstow et al. 1995a) and RE J0720-318
(Barstow et al. 1995b, Vennes and Thorstensen 1994). All three lie in
close, post-common envelope binaries with cool red dwarf companions,
have masses consistent with post-AGB evolution, and 
represent the fifth of Bergeron et al.'s DAO subgroups. 
Burleigh and Barstow (1995) showed that the failure to detect any
further DAOs in the EUV was either because the stars
lay in high interstellar hydrogen column directions, or because these hot
objects must 
have substantial quantities of elements heavier than He levitated in their 
atmospheres, providing the opacity to block any EUV flux.
In addition, given that DAs outnumber DAOs by about
a hundred to one in white dwarf catalogues (e.g. McCook and Sion 1987),
  and there are about 110 DAs identified in the
ROSAT Wide Field Camera (WFC) survey  (e.g. Pye et al. 1995), we should
not necessarily expect to see any further examples in the EUV.

The optical spectrum of RE J0720$-$318 (Figure 1) clearly shows the
distinctive HeII absorption feature at 4686\AA,  
which can only be fitted with a homogeneous H+He 
atmospheric stucture. From modelling this spectrum, Barstow et al.
(1995b) derive the white dwarf parameters T$_{eff}$, log g and log He/H
given in Table 1.
The binary nature of the system is easily recognized from the 
variable narrow emission lines in the cores of the Balmer absorption dips. 
These arise from reprocessing of the EUV radiation from the white dwarf
on the surface of the red dwarf companion. RE J0720$-$318 has the lowest
helium abundance of any of the DAOs studied by Bergeron et al. (1994)
and Burleigh and Barstow (1995).   
Tweedy et al. (1993) and Barstow et al.
1995b) speculate that the helium in these stars may result from accretion
from the stellar wind of the M dwarf companion.

\begin{table}
\caption{White dwarf parameters (from optical fit)}
\begin{tabular}{ccccccccc}
\hline
Parameter & Value & 1$\sigma$ limits \\
\hline
Temp. (K) & 53630 & 52400$-$54525 \\
log g     & 7.64 & 7.54$-$7.74 \\
log He/H  & -3.56 & -3.43$-$-3.70 \\
M (M$_\odot$) & 0.55 & $\pm$0.03 \\
\hline
\end{tabular}
\end{table}

\begin{figure}
\vspace*{5.1cm}
\includegraphics{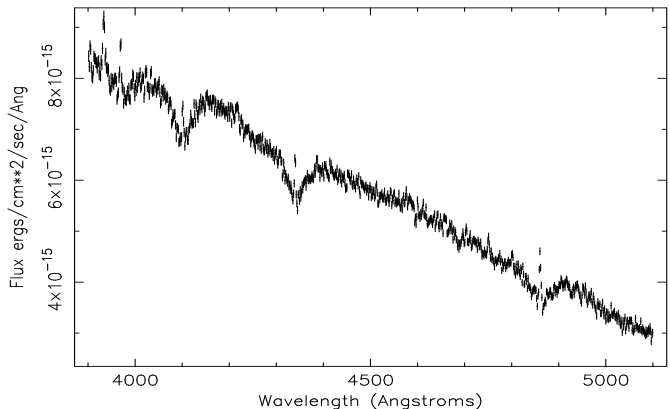}
\caption{Co-added optical spectrum of RE J0720$-$318, showing the weak
HeII 4686{\AA} feature and the characteristic narrow emission lines}
\end{figure}

\section{EUVE and IUE Observations and data reduction}

RE J0720$-$318 was observed by EUVE in `dither' mode between 1995
December 16-20, and was detected in all three wavebands with exposure
times of 122,459s (SW, 70$-$190{\AA}), 121,067s (MW, 140$-$380{\AA}) 
and 112,861s (LW, 280$-$760{\AA}). The dither mode
consists of a series of pointings slightly offset in different directions
from the nominal source position. Systematic
variations in the EUVE detector efficiency originally limited the
signal-to-noise that could be achieved to $\approx$5. By using the dither
mode to average out flat field variations, a substantial fraction of this 
fixed pattern noise can be removed. 
This technique has improved the S/N of the dithered spectrum
of HZ43 by around a factor 4 (Dupuis et al. 1995, Barstow, Holberg and
Koester 1995).  
Therefore, to take  any residual fixed pattern noise into account, we
quadratically add a 5\% systematic to the statistical errors on the data.
We have extracted the spectra from the images ourselves, using standard
IRAF procedures. Our general reduction techniques are described in more
detail in our  earlier work (e.g. Barstow, Holberg and Koester 1994).

Two low resolution IUE SWP spectra 
(SWP54496 and SWP54497, exposure $=$ 40 minutes
each) were obtained by us on 1995 April 24. These have been extracted and
calibrated with the NEWSIPS processing, which gives an absolute error on
each data point, and then co-added to improve the
overall signal-to-noise (Figure 2).
 
We have also included the ROSAT PSPC soft band (0.1-0.4keV) 
X-ray count rate (0.455 counts/sec) in our analysis (Fleming et al. 1996). 
Significantly, 
RE J0720$-$318 was not detected in the PSPC hard (0.4-2.4keV) band. We
can assume, therefore, that the vast majority of the EUV and soft X-ray
flux comes from the white dwarf alone, and that the red dwarf companion
is not active.

\section{Analysis and Results}

We have fitted a set of fully line blanketed 
stratified and homogeneous H$+$He models,
computed by Koester (1991), to the far-UV and EUV data,  using the XSPEC
spectral fitting programme. We have described the detailed use of XSPEC
for analysis of optical, UV and EUV data in several previous publications
(e.g. Barstow, Holberg and Koester 1994). 
The stratified model assumes plane parallel
geometry with a thin H layer overlying a helium atmosphere, under LTE
conditions. The H layer mass is covered in the approximate range
10$^{-16}$$>$M$_{H\odot}$$>$10$^{-11}$. The second model structure assumes
a homogeneous distribution of H$+$He, also under LTE conditions, in
the range $-$8$>$log He/H$>$$-$2. 
When comparing the IUE and EUVE data sets with the models, we utilise the V
magnitude (V=14.87$\pm$0.04) as a normalisation point, 
 calculated by Barstow et al. (1995b) to
account for the contribution by the red dwarf companion.

\subsection{Far-UV data}

The co-added low resolution IUE SWP spectrum (Figure 2) clearly shows no 
evidence for 
weak HeII absorption at 1640{\AA} or any other heavy element features. 
This is in contrast to  far-UV 
spectra of the similar system RE J1016$-$053. 
Tweedy et al. (1993) report weak HeII and CIV absorption lines in
this star at 1640{\AA} and 1549{\AA} respectively, and Barstow et
al. (1995a) also report a CIV 1549{\AA} feature in the other ROSAT pre-CV
DAO/dM binary RE J2013$+$400. However, they do not see a HeII line at 
1640{\AA}. The upper limit to the helium content of RE J2013$+$400 from
the IUE spectrum is log He/H=-2.5, entirely consistent with the optical
data.

\begin{figure}
\vspace*{5.1cm}
\includegraphics{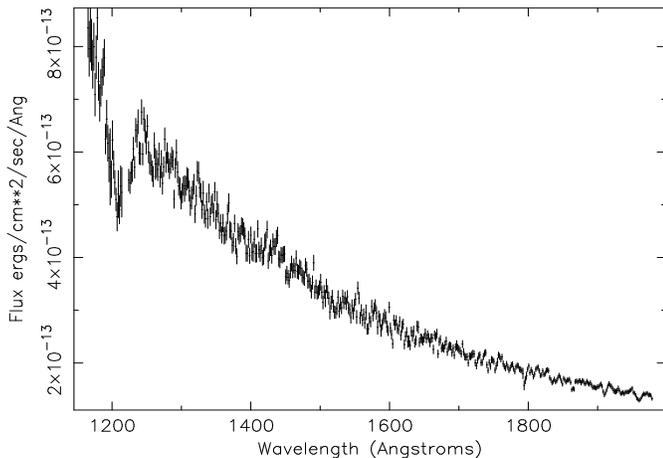}
\caption{Co-added IUE SWP Spectrum of RE J0720$-$318. Note the absence of
any weak HeII absorption feature at 1640\AA}
\end{figure}

For RE J0720$-$318, assuming a homogeneous model structure and fitting 
the optically derived
T and log g, we find that the helium content implied from a fit to the
optical data (log He/H=-3.56) can easily be hidden within the general
scatter of the data points around 1640{\AA} in the IUE spectrum. Indeed,
a helium line would only be visible above the noise for log
He/H$>$-2.5.   

\subsection{The EUVE spectrum}

The entire EUVE spectrum is shown in Figures 3 and 4. The most striking
feature is the saturated HeI absorption edge at 504{\AA}, dominating the long
wavelength spectrum. Nothing like this
has ever been seen in any other EUVE spectra of white dwarfs (e.g. Dupuis
et al. 1995, Barstow, Dobbie and Holberg 1996). 
In fitting the EUVE spectrum and the PSPC data point, we utilise the
optically determined T and log g (Barstow et al. 1995b, see
Table 1), and allow these only to vary between their 1$\sigma$
confidence limits. 
Figure 3 shows the best fit using a homogeneous model. Although the
HeI 504{\AA} edge is well matched, the predicted HeII edge at 228{\AA} 
and the short wavelength continuum do not match the observed fluxes. 

The best fit stratified model is shown in Figure 4. All the major
features are accurately reproduced, in particular the HeII series of absorption dips
converging on 228{\AA}, the 206{\AA} feature, the sw spectrum and 
 the HeI 504{\AA} edge. The best fit parameters are listed in Table 2.

\begin{figure}
\vspace*{5.1cm}
\includegraphics{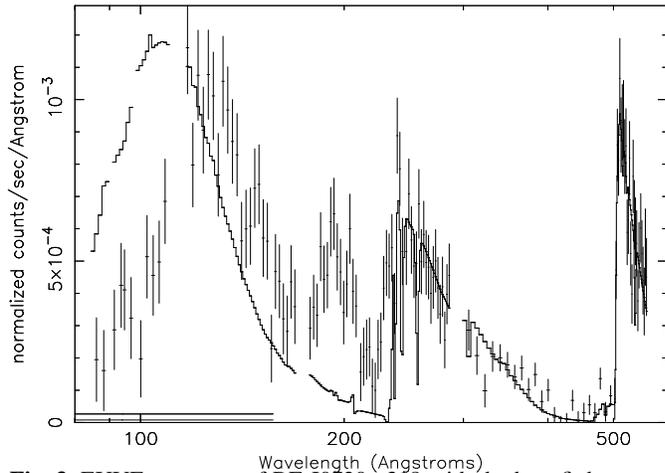}
\caption{EUVE spectrum of RE J0720$-$318 with the best fit homogeneous
model (every 2nd data point has been removed for clarity)}
\end{figure}

\begin{figure}
\vspace*{4.6cm}
\includegraphics{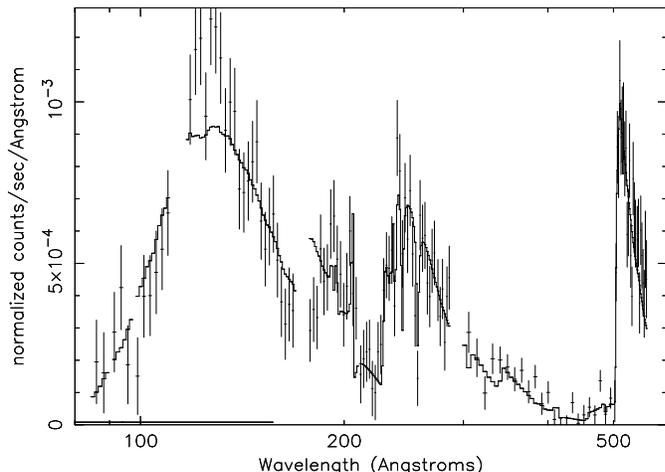}
\caption{EUVE spectrum of RE J0720$-$318 with the best fit stratified
model (every 2nd data point has been removed for clarity)}
\end{figure}

\begin{table}
\caption{Interstellar H and He column densities and H layer mass derived 
from the best fit stratified model}
\begin{tabular}{ccccccccc}
\hline
Parameter & Value & 1$\sigma$ limits \\
\hline
N$_H$ & 2.30$\times$10$^{18}$ & 2.25$-$2.36$\times$10$^{18}$\\
N$_{HeI}$ & 1.44$\times$10$^{18}$ & 1.41$-$1.47$\times$10$^{18}$\\
N$_{HeII}$ & 9.78$\times$10$^{17}$ & 8.58$-$10.03$\times$10$^{18}$\\
M$_H$ & 3.07$\times$10$^{-14}$M$_{\odot}$ &
2.93$-$3.21$\times$10$^{-14}$M$_{\odot}$\\
%4$\times$10$^{19}$ & 7.0 & 36600 & 0.032 & 0.38 & 267 & 40 & 59 & 266 \\ 
                                          \hline
\end{tabular}
\end{table}

\section{Discussion}

The EUVE spectrum of RE J0720$-$318 has yielded three important results.
Firstly, 
in contrast to the optical spectrum, which can only be matched with a
mixed H$+$He model structure, we have shown that the EUV spectrum 
 can only be reproduced by a stratified atmosphere. Secondly, we find the
white dwarf has an unexpectedly thin H layer, 3$\times$10$^{-14}$M$_\odot$.
Finally, there appears, compared with HI, 
to be an unusually high line-of-sight interstellar helium column
density to the system (see Table 2). 
The implied hydrogen ionization fraction, 
assuming a cosmic He/H ratio, is 90\%, and the helium ionization
fraction is 40\%. 

The failure to model the EUVE spectrum with a homogeneous atmosphere
confirms the result of Barstow et al. (1995b), who could not match the
ROSAT WFC and PSPC data points with a mixed H/He structure. 
These authors also failed to fit the ROSAT data with a stratified model.
However, the EUVE spectrum reveals line of sight absorbing material with
a HeI/HI ratio far removed from the canonical (cosmic) value of 0.1
originally assumed, and an additional HeII component not previously taken
into account.  
We have now shown that the EUV and soft X-ray data can indeed be matched 
by a stratified model, although this model cannot reproduce the HeII
4686{\AA} feature in the optical region. 
This has important implications for our understanding of the white
dwarf's structure and evolution, and for the binary itself. 

Since the EUV radiation originates from hotter, deeper layers of the
white dwarf atmosphere, the underlying structure of the star must be
stratified. 
The helium present in the optical spectrum 
may reside on the surface of the white dwarf in a thin mixed
layer. This would suggest that the helium is being accreted from
the red dwarf companion via a stellar wind, and is not intrinsic to the
white dwarf itself. 

The hydrogen layer mass derived from the stratified fit is 
the lowest found for any white dwarf from
EUVE spectra (Barstow, Dobbie and Holberg 1996). 
If this had been an isolated star it would probably have evolved into a
normal DA, but as it is in a close binary system the thin H layer 
may well be the result of mass transfer between the two
components during an earlier common envelope phase. 
Thereafter it is possible for any moderate mass loss to bring some of the 
underlying helium to the surface, giving the optical HeII 4686{\AA} line. 
However, analysis of the fit to the medium waveband spectrum
(Figure 4, inset) clearly shows that while the HeII series coverging on
228{\AA} is photospheric in origin, the edge itself must be due to
cooler interstellar or possibly circumstellar material. 

The HeI/HI ratio of $\sim$1 is, by around a factor ten, the highest
measured in any direction in the sky from EUVE data 
(e.g. Barstow, Dobbie and Holberg  
1996). This is so high that, taken together with the $\sim$90\% hydrogen 
ionization fraction (also the highest measured from EUVE data), 
we must question whether the absorbing gas lies 
in the local interstellar medium between
ourselves and  RE J0720$-$318, or if in fact this gas lies in the
vicinity of the binary system. 

It is possible that most of the
helium resides in the RE J0720$-$318 system itself, in the form of a
circumbinary gas. 
Recent theoretical studies of CE evolution by Terman and
Taam (1996) suggest that circumbinary disks are likely to form in post-CE
systems. The spiral-in process decelerates so rapidly 
in the final stages of CE evolution that material in the immediate
vicinity of the binary cores is not expected to be ejected from the
system. Instead, this forms a disk in the orbital plane of the binary 
(since studies show that ejection of the CE mainly takes place in this
direction). It should be noted that Barstow et al . (1995b) find evidence
to suggest  the orbital inclination of the binary may be as high as
85$^\circ$, and thus it is possible that we are seeing the system through
an optically-thin disk. 
The presence of this disk may affect the orbital evolution of
the detached pre-CV, as is known to occur in pre-main sequence stars,  
through tidal and resonant interaction between the binary and disk.
Processes such as gravitational radiation and magnetic braking have often
been discussed as mechanisms in the evolution of pre-CV binaries, but
thus far the effect of a circumbinary disk has been generally overlooked.
The possible detection of such a disk in a post-CE system therefore has
profound implication for our understanding of the formation and evolution
of close binaries.

\begin{acknowledgements}

MRB, MAB and PDD acknowledge the support of PPARC, UK. We wish to thank
Detlev Koester for the use of his white dwarf model atmosphere grids, and 
Jay Holberg and Jim Collins at the University of Arizona for their help
in obtaining and reducing the IUE data.

\end{acknowledgements}

\end{document}